\RequirePackage[2020-02-02]{latexrelease}
\documentclass[preprint,aps,floatfix]{revtex4}
\DeclareUnicodeCharacter{00A0}{ }
\usepackage{graphicx}
\usepackage{textgreek}
\usepackage{verbatim} 
\usepackage{amssymb}
\usepackage{url}
\usepackage{amsmath,amssymb}
\usepackage{mathtools} 
\usepackage[symbol]{footmisc}
\usepackage{subcaption}
\usepackage{capt-of}
\usepackage{bigints}
\usepackage{romannum}
\usepackage{bm}
\usepackage{xcolor} 
\usepackage{soul} 
\AtBeginDocument{\pagenumbering{arabic}}

\def\be{\begin{eqnarray}}
\def\ee{\end{eqnarray}}

\begin{document}

\title{Role of Coulomb-nuclear breakup of $^{6,7}$Li projectiles with heavy deformed $^{232}$Th target}

  \author{D. Patel${}^1$\footnote{dipikapatel@phy.svnit.ac.in(corresponding author)}, J. Rangel${}^2$\footnote{jeannierangel@gmail.com}, and J. Lubian${}^3$\footnote{jlubian@id.uff.br}}
  \affiliation{${}^1$ Department of Physics, Sardar Vallabhbhai National Institute of Technology, Surat, 395 007, India.\\ ${}^2$ Departamento de Matemática, Física e Computação, Universidade do Estado do Rio de Janeiro, Faculdade de Tecnologia, 27537-000 Resende, Rio de Janeiro, Brazil.\\ ${}^3$ Instituto de Física, Universidade Federal Fluminense, Gragoatá, Niterói, RJ 24210-340, Brazil.}
  
\begin{abstract}
The significance of both Coulomb and nuclear couplings and their interference effects in the breakup processes of $^{6,7}$Li with a non-spherical nucleus $^{232}$Th has been evaluated. The continuum discretized coupled channel(CDCC) calculations are carried out in a nonstandard way, using short-range imaginary potentials for the fragment-target interaction at energies close to the Coulomb barrier. The present calculations employing short-range imaginary potentials exhibit better agreement with the experimental elastic scattering angular distributions than those using standard systematic value ($0.78\times W_{SPP}$) used to describe elastic scattering. Including the excitation of the $^{232}$Th inelastic shows significant coupling effects on the elastic scattering below the barrier energies compared to higher incident energies. Subsequently, the CDCC framework was used to analyze the nuclear, Coulomb, and total breakup predictions separately. The breakup cross sections for the $^{6}$Li+$^{232}$Th system are greater than those for the $^{7}$Li+$^{232}$Th system across various energies. The present study predicts destructive Coulomb-nuclear interference in the breakup processes involving both $^{6}$Li and $^{7}$Li projectile nuclei with the deformed $^{232}$Th target. Additionally, the breakup reaction cross-sections are compared with experimentally measured fusion cross-sections near the barrier energies for both $^{6,7}$Li+$^{232}$Th systems.
\end{abstract}
\maketitle
\section{Introduction}
In recent years, there has been significant interest in studying heavy ion reactions involving projectile nuclei near the drip-line, such as $^{6}$He, $^{9,11}$Li, $^{10,11,14}$Be, ${etc}$ \cite{Canto2006,Torabi:2020gie,Back:2014ypa}, aimed at probing the internal structure of nuclei by investigating the effects of breakup coupling on various reaction channels. However, due to the challenges posed by limited beam intensity and the availability of unstable weakly bound nuclei, experiments have focused extensively on stable weakly bound nuclei like $^{6,7}$Li and $^{9}$Be. The structural similarities between stable and unstable weakly bound nuclei have further underscored the importance of studying reactions involving stable isotopes.

Both experimentally and theoretically, it has been observed that in reactions involving weakly bound nuclei, the breakup channel does not disappear near the Coulomb barrier; instead, it significantly increases in magnitude because the projectile does not need to tunnel the Coulomb barrier to excite the continuum states (populated in the
breakup process). This phenomenon results in the emergence of a repulsive polarization potential, a well-established observation known as the Breakup Threshold Anomaly (BTA)\cite{Gomes:2005hb,Chamon:2005sa,Patel:2014pja,Patel:2015ava}.
The elastic scatterings involving $^{6,7}$Li projectiles are influenced by the continuum states, owing to the low thresholds. Hence, it is necessary to include the continuum in the coupled channel calculation, which can be done with the CDCC method.

The theoretical framework for describing elastic scattering is relatively well-established, even for two-neutron halo nuclei \cite{Matsumoto:2004ck,Fernandez-Garcia:2015gra}. However, the interplay between competing Coulomb and nuclear breakup processes is more complex and has been the focus of extensive investigation by various researchers\cite{Rangel:2016kqs,Marta:2008zz,Dasso:1999zz,Kucuk:2012gd,Nunes:1999rx,Tostevin:2000jd,Goldstein:2006sv,Crespo:2011zza,Keeley:2010zza,Rusek2009}. These effects are crucial for understanding reaction dynamics in nuclear astrophysics, heavy-ion collisions, and near-barrier fusion reactions. The Coulomb and nuclear effects and their interference in breakup reactions have garnered significant interest in recent studies \cite{Dasso:1996eyc,Hussein:2006yj,Patel:2017mio,Kumar:2012yp}. Most of these studies have been performed on systems involving spherical targets to isolate the breakup mechanism.

In this work, a detailed analysis has been performed using the Continuum Discretized Coupled Channel (CDCC) method to investigate the interactions of $^{6,7}$Li projectiles with the heavy deformed target $^{232}$Th to verify whether the conclusions taken for systems involving spherical targets about the observed Coulomb-nuclear interference remain valid for a system involving a deformed target. Given that $^{232}$Th is a deformed actinide nucleus with a notable quadrupole deformation parameter $\beta_{2}$=0.261~\cite{Raman:2001nnq}, the effects of couplings on its inelastic states have also been examined through Coupled Channel calculations. The individual contributions of Coulomb and nuclear breakup couplings have been analyzed for both the $^{6,7}$Li+$^{232}$Th systems.
To conduct these breakup calculations, two different approaches have been employed for fragment-target potentials. Furthermore, we have investigated the interference effects between Coulomb and nuclear breakup couplings on elastic and direct breakup reactions by computing breakup cross-sections separately. The cross-sections have been evaluated at various energies near the Coulomb barrier, and detailed angular distributions are presented to explore the complexities of Coulomb-nuclear interference phenomena. All theoretical calculations performed in this work were done with the FRESCO code\cite{Thompson:1988zz}.
\section{Coupled channel calculations for $^{232}$Th inelastic excitations}
\begin{figure*}
\center
\centering
\resizebox{0.8\textwidth}{!}{%
  \includegraphics{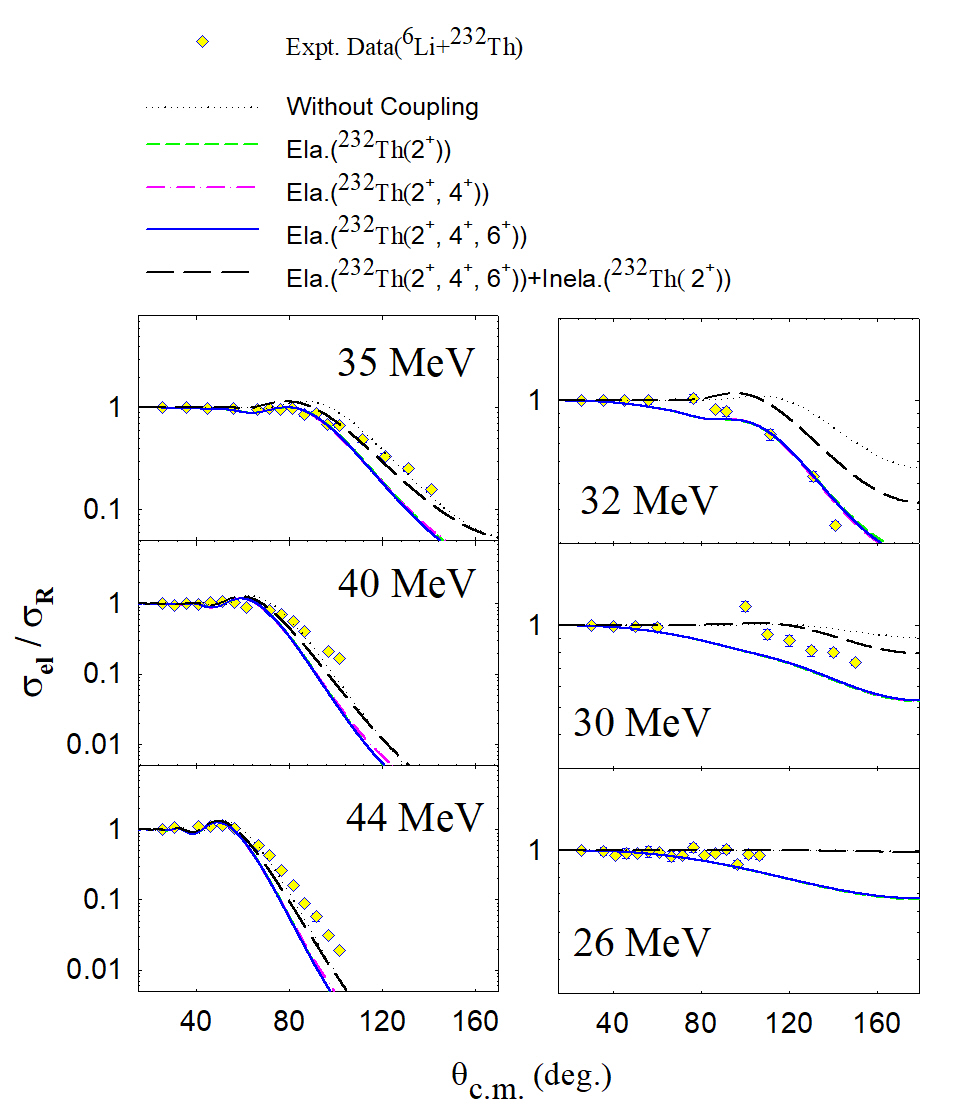}
}
\caption{(Color online)Quasi-elastic scattering data for the $^{6}$Li+$^{232}$Th system \cite{Dubey:2014ota} is compared with the CC results at 44, 40, 35, 32, 30, and 26MeV(refer to the text for detailed discussion).}
\label{fig:CC_ela+inela-6lith.eps}       
\endcenter
\end{figure*}

\begin{figure*}
\center
\resizebox{0.8\textwidth}{!}{%
  \includegraphics{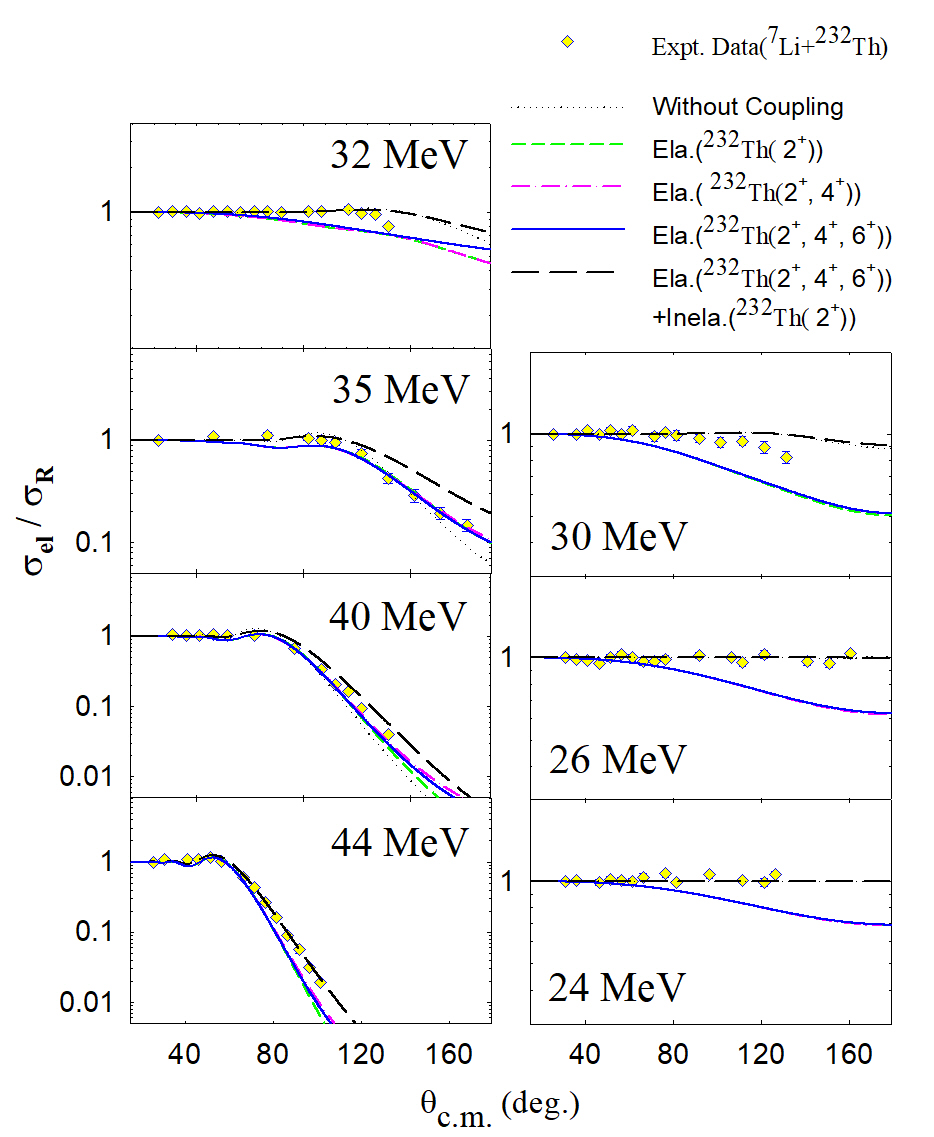}
}
\caption{(Color online)Quasi-elastic scattering data for the $^{7}$Li+$^{232}$Th system \cite{Dubey:2014ota} is compared with the CC results at 44, 40, 35, 32, 30, 26, and 24MeV(refer to the text for detailed discussion).}
\label{fig:CC_ela+inela-7lith.eps}       
\endcenter
\end{figure*}

The coupled channel approach investigates the impact of inelastic excitations in $^{232}$Th on the quasi-elastic scattering angular distributions. 
The ground-state rotational band of $^{232}$Th, including the excited states 2$^{+}$(0.049MeV), 4$^{+}$(0.162MeV),
6$^{+}$(0.3332MeV), is included in the calculations. A quadrupole deformation parameter of $\beta_{2}$=0.261~\cite{Raman:2001nnq} is used to derive the coupling potential, considering both Coulomb and nuclear interactions due to $^{232}$Th inelastic excitations. The calculations were performed with the S$\tilde{a}$o Paulo Potential (SPP) ~\cite{Thompson:1988zz} for the real part, while short-range imaginary potentials with parameters $W_{i}$=50 $MeV$, $r_{i}$=1.06 $fm$, and $a_{i}$=0.2$fm$ were employed. Keeping the same potentials, the calculations were carried out for several energies at around the Coulomb barrier energies for both $^{6,7}$Li+$^{232}$Th systems. The results of the coupling of inelastic states for these calculations are compared with the experimental angular distributions \cite{Dubey:2014ota} for eight $^7$Li projectile energies (24, 26, 30, 32, 35, 40, and 44 MeV) and six $^6$Li energies (26, 30, 32, 35, 40, and 44 MeV),
as shown in Figs. \ref{fig:CC_ela+inela-6lith.eps} and \ref{fig:CC_ela+inela-7lith.eps}. 

In these figures, dashed,  dashed-dotted, and solid lines show results for the elastic scattering angular distributions, including states in the coupled scheme, from 2$^{+}$ up to 6$^{+}$, respectively. 
The potential scattering results are shown by dotted lines, which means only the ground state is considered. One can observe that at higher incident energies, the inelastic state coupling is small; however, as incident energies decrease, the inelastic state coupling appears to be more significant. It is important to note that the experimental data are of quasi-elastic nature, as isolating the lowest energy state, 2$^{+}$(0.049MeV) of $^{232}$Th, is challenging. A more accurate representation of the experimental data was achieved only after incorporating the inelastic contributions of the 2$^{+}$(0.049MeV) state into the elastic component (long-dashed line in these figures). 

However, the remaining discrepancy between the predictions and the experimental data, even after including the 2$^{+}$ inelastic state, can be attributed to breakup processes, which play a significant role in the interactions of the weakly bound $^{6,7}$Li nuclei. First, it is well known that the static effects of the cluster model for this nucleus hinder the Coulomb barrier, and the flux from the elastic channel is lost to the fusion mechanism and should be very important at low energies. To better understand the overall estimations, it is important to do complete coupled channel calculations, including breakup channels besides target excitations. Unfortunately, it is not well-established numerically. The observed hindrance below the Coulomb barrier ($\sim$32MeV in the laboratory frame) may also be because of the coupling to transfer channels, which may take flux from the elastic channel \cite{Keeley:2001zu}. Also, the effect of the transfer channels at energies below the Coulomb barrier is well-known for the systems that involve the spherical target\cite{Luong:2013cga,DiPietro:2013ama}.

\begin{figure*}
\center
\resizebox{0.8\textwidth}{!}{%
  \includegraphics{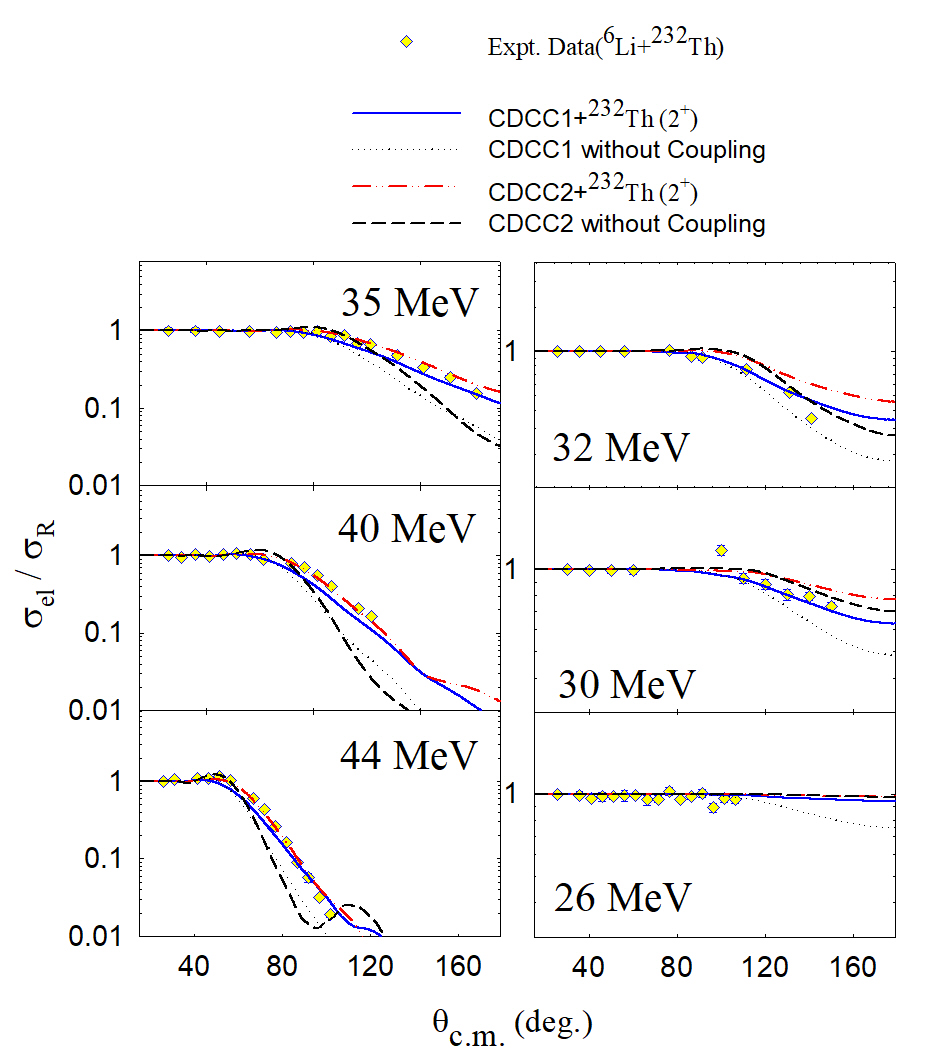}
}
\caption{(Color online)Quasi-elastic scattering angular distributions at 44, 40, 35, 32, 30, and 26MeV for the $^{6}$Li+$^{232}$Th system in comparison with data \cite{Dubey:2014ota}. The lines are the results of CDCC calculations. (refer to the text for detailed discussion).}
\label{fig:comp-0.78-50img-6lith.eps}       
\endcenter
\end{figure*}

\begin{figure*}
\center
\resizebox{0.8\textwidth}{!}{%
  \includegraphics{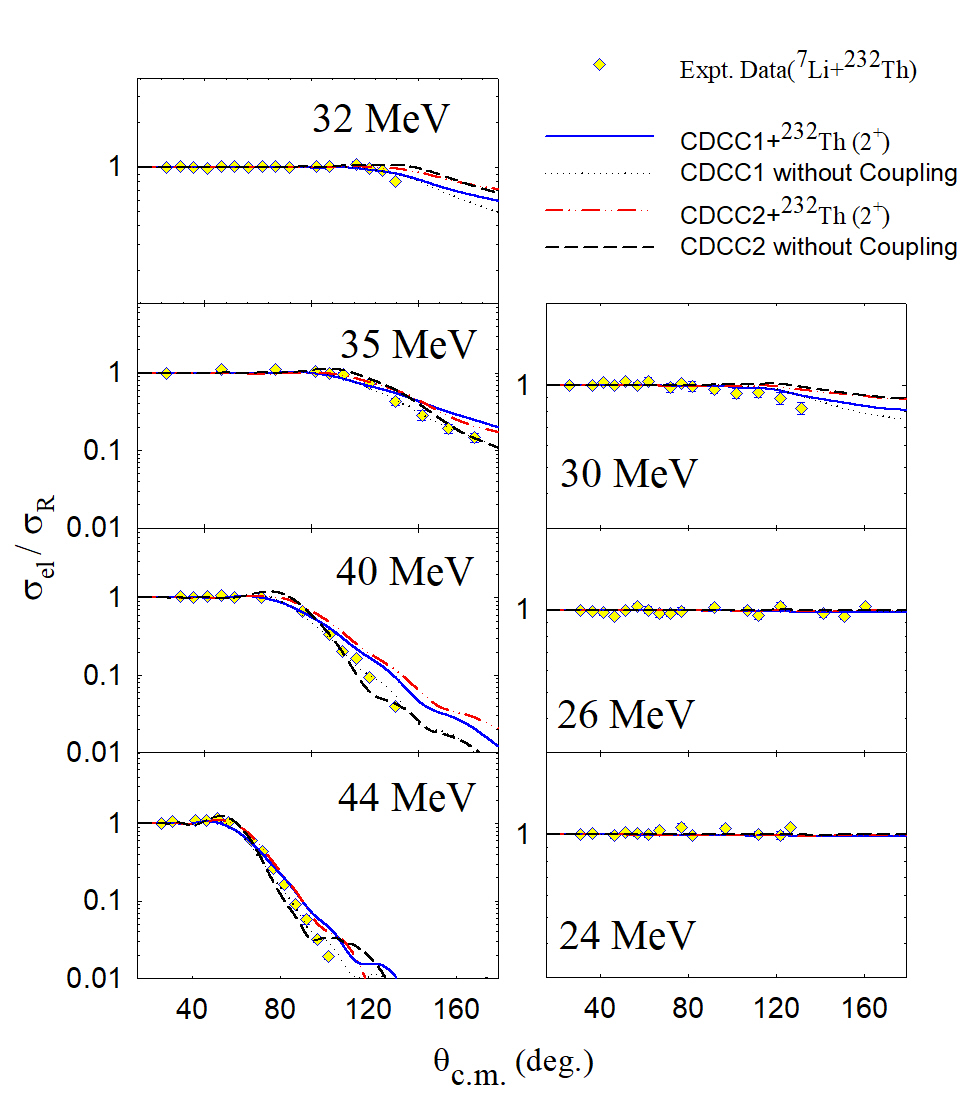}
}
\caption{(Color online)Quasi-elastic scattering angular distributions at 44, 40, 35, 32, 30, 26, and 24MeV for the $^{7}$Li+$^{232}$Th system in comparison with data \cite{Dubey:2014ota}. The lines are the results of CDCC calculations. (refer to the text for detailed discussion).}
\label{fig:comp-0.78-50img-7lith.eps}       
\endcenter
\end{figure*}
\section{Continuum Discretized Coupled channel calculation for $^{6,7}$Li+$^{232}$Th reactions}
 Breakup effects of weakly bound nuclei on elastic scattering are well addressed by the CDCC method. This method was introduced in the seventies to analyze deuteron scattering on heavy targets \cite{Rawitscher:1974zz,Sakuragi:1986vq}. The CDCC method accurately approximates the three-body wave function and is well-adapted to breakup reactions. $^{6,7}$Li are treated as a two-cluster system of core-fragment, where $\alpha$ is considered as core and $^{2}$H and $^{3}$H as fragments with separation energies of $\sim$1.47MeV and $\sim$ 2.45MeV, respectively. For both the $^{6,7}$Li+$^{232}$Th systems, the multipole expansion of the potentials was carried out considering multipoles up to $\lambda$=4. The adopted CDCC binning method includes a matching radius of $r_{bin}$=40$fm$ to guarantee negligible transition matrix elements between the bin states at larger distances. In the solution of the coupled channel equations, a relative angular momentum up to $J$=1000$\hbar$ was considered, and couplings up to $R_{coup}$= 800$fm$. The binning scheme technique for the continuum spectrum of $^{6,7}$Li projectile nuclei is implemented in the same manner as outlined in the CDCC calculations of Ref. \cite{Cortes:2020rak}.

\begin{table}
\center
\caption{The potential depths, radii, and diffusenesses of Ground State(G.S.), Resonant State(R.S.), Bound States(B.S.) of $^{6,7}$Li projectiles. The potential depths are given in MeV, and the radii and diffusenesses in fm\cite{Diaz-Torres:2003usc}. }
\label{tab:1}       
\begin{tabular}{lcccccc}
\hline\noalign{\smallskip}
projectiles & $V_{0}$ & $r_{0}$ & $a_{0}$ & $V_{0}^{s.o.}$ & $r_{0}^{s.o.}$ & $a_{0}^{s.o.}$  \\
and states\\
\noalign{\smallskip}\hline\noalign{\smallskip}
$^{6}$Li(G.S)  & -78.46 & 1.15 & 0.7 & & & \\
$^{6}$Li(R.S.) & -80.0 & 1.15 & 0.7 & 2.25 & 1.15 & 0.7  \\
$^{7}$Li(B.S.)& -108.1  & 1.15 & 0.7 & 0.9875 & 1.15 & 0.7  \\
$^{7}$Li(R.S.) & -109.89 & 1.15 & 0.7 & 1.6122 & 1.15 & 0.7  \\
\noalign{\smallskip}\hline
\end{tabular}
\endcenter
\end{table}

\begin{table}
\center
\caption{Experimental\cite{Diaz-Torres:2003usc} and Theoretical(Present calculation) Resonant state energies and widths of the $^{6,7}$Li projectile nuclei}
\label{tab:2}       
\begin{tabular}{lcccccc}
\hline\noalign{\smallskip}
R.S.($^{6,7}$Li) & $l$ & $E_{R.S.}^{Theor.}$ &  $\Gamma_{R.S.}^{Theor.}$ & $E_{R.S.}^{Expt.}$ & $\Gamma_{R.S.}^{Expt.}$ \\
and $j^{\pi}$ \\
\noalign{\smallskip}\hline\noalign{\smallskip}
$^{6}$Li(3$^{+}$) & 2 & 0.717 & 0.02 & 0.716 & 0.024 \\
$^{6}$Li(2$^{+}$) & 2 & 3.14 & 1.88 & 2.84 & 1.7 \\
$^{6}$Li(1$^{+}$) & 2 & 4.06 & 3.5 & 4.18 & 1.5 \\
$^{7}$Li(7/2$^{-})$ & 3 & 2.15 & 0.1 & 2.16 & 0.093 \\
$^{7}$Li(5/2$^{-}$) & 3 & 4.54 & 0.88 & 4.21 & 0.88 \\
\noalign{\smallskip}\hline
\end{tabular}
\endcenter
\end{table}
In this approach, the continuum is discretized in energy space using bins of constant width. For the resonant part of the continuum, narrower bin widths are used, while larger bins are used as the energy approaches the maximum cutoff energy. The potential used to generate the bin wave packets (for resonant and nonresonant states), resonant state energies, and widths utilized in the present calculations for both $^{6,7}$Li projectile nuclei are listed in Tables \ref{tab:1} and \ref{tab:2}. The maximum energy for the continuum states was set to 10~\text{MeV} for $^{6}$Li and 11~\text{MeV} for $^{7}$Li. The interaction between the core nuclei ($\alpha$) and the valence ($d,t$) particles in the clusters of $^{6,7}$Li is described using a Woods-Saxon potential to model the ground state and the unbound resonant states~\cite{Cortes:2020rak,Diaz-Torres:2003usc,Otomar:2009zz}. 

To conduct a systematic investigation, we first applied the standard method(CDCC1), utilizing the double-folding São Paulo Potential (SPP)~\cite{Thompson:1988zz} for the real part of the potentials governing the interactions of $\alpha,d$+$^{232}$Th and $\alpha,t$+$^{232}$Th. For the imaginary part of the optical potentials of the fragment-target interaction, the SPP systematic was used. That is, a scaling factor of 0.78 was used. A widely adopted approach is to scale the imaginary part, using the real double folding potential, by a factor, often set to 0.78, to account for channels weakly coupled to the elastic~\cite{Matsumoto:2004ck,Joshi:2022xyl,Linares:2021cdr,Pakou:2022bwg}. We emphasize that this systematic method was introduced to describe heavy-ion elastic scattering, as the system does not have a strong coupling of the elastic and other reaction channels. Clearly, this is not the case for the reaction of the fragments and the target due to the strong collectivity of the low-lying states of the $^{232}$Th target. This would correspond to the standard CDCC calculations where fragment-target optical potentials are introduced in a way that describes their elastic scattering.

A separate set of CDCC calculations (CDCC2) was performed using a nonstandard approach, incorporating short-range imaginary potentials. In these calculations, the real part was retained from the SPP, while a short-range imaginary potential was employed for the interactions of $\alpha,d$+$^{232}$Th and $\alpha,t$+$^{232}$Th. These potentials were described by Woods-Saxon functions with a depth $W_{i}=50$~MeV, the radial parameter $r_{i}=1.06$~fm, and diffusivity $a_{i}=0.2$~fm. 


Performing CDCC calculations that simultaneously incorporate both projectile breakup and target excitation remains an open challenge. To compare the results for the elastic scattering from CDCC calculations with the quasi-elastic angular distributions, one needs to add the cross section of the unresolved first 2$^{+}$(0.049MeV) excited state of the $^{232}$Th target. In the previous section, we showed that the contribution of this unresolved state is relevant. So, we added the CC cross section of the previous section to the CDCC elastic angular distribution to compare with the experimental data. The angular distributions will be called CDCC1+$^{232}$Th (2$^+$) and CDCC2+$^{232}$Th (2$^+$).

The corresponding results are presented in Fig. \ref{fig:comp-0.78-50img-6lith.eps} and Fig. \ref{fig:comp-0.78-50img-7lith.eps} at various energies. The solid and dash-dotted lines in these figures represent the CDCC results obtained using the standard method (CDCC1) and the short-range imaginary potential(CDCC2), respectively. Meanwhile, the dotted and dashed lines correspond to the results without any coupling effects, using the standard method(CDCC1) and the short-range imaginary potential(CDCC2), respectively.

Interestingly, using a short-range imaginary potential provided a slightly better description than the standard method at energies above the Coulomb barrier. This observation arises because standard CDCC calculations require optical potentials for the interaction between the projectile fragments and the target to include an imaginary component extending to the surface, which effectively describes the elastic scattering of each fragment with the target. To our knowledge, such a potential is not currently available in the literature, nor is there experimental data to constrain the potential parameters.

Another noteworthy point is that breakup channels are known to produce repulsive polarization potential, while the coupling to collective states is attractive. The polarizations of opposite signs might cancel at energies above the Coulomb barrier, and this could be the reason for having good results with internal imaginary potentials that correspond to the case where only the fusion mechanism is relevant. This kind of suggestion needs theoretical support in the future.

\section{Nuclear, Coulomb, and Interference Constituents by CDCC calculations}

The nuclear, Coulomb, and interference constituents in the elastic scattering and breakup angular distributions have been obtained at various incident energies for both $^{6,7}$Li projectiles implementing the standard CDCC method as described in the previous section. In this work, we separate the diagonal and off-diagonal components of the full projectile-target interaction in channel space, as shown below:
\begin{equation}
U = U_{opt} + \Delta V,
\end{equation}
It is important to note that  $U_{opt}$  serves as the optical potential, while $\Delta V$ represents the channel coupling interaction. Furthermore, the coupling potential is divided into its Coulomb and nuclear components.
The coupling potential $\Delta V$ was further divided into Coulomb and nuclear components:
\begin{equation}
\Delta V = \Delta V{(Coul.)} + \Delta V{(Nucl.)}.
\end{equation}
The breakup cross-sections were analyzed in three distinct steps: (I)Using the full potential ($U$) as shown in eq. (1), (II)Using $U_{opt}+\Delta V{(Coul.)}$ to isolate the Coulomb breakup cross-section ($\sigma_{CBU}$) and (III)Using $U_{opt}+$ $\Delta V{(Nucl.)}$ to determine the nuclear breakup cross section ($\sigma_{NBU}$). 
The Coulomb-nuclear interference part was obtained by subtracting Coulomb and Nuclear breakup cross-sections from the total cross-sections ($\sigma_{TBU}$ - $(\sigma_{CBU}$ + $\sigma_{NBU}$)). As mentioned in the introduction, we emphasize that this is an approximation usually used for that kind of study.

The integrated breakup cross-sections at various energies, for Nuclear and Coulomb, have been obtained from the calculated breakup angular distributions. Tables \ref{tab:3} and \ref{tab:4} describe the separated Coulomb and nuclear components and their interference effect at various incident energies for both $^{6,7}$Li+$^{232}$Th systems. It can be seen that breakup cross-sections increase as projectile energy increases for both $^{6,7}$Li+$^{232}$Th systems. 
\begin{figure*}
\center
\resizebox{0.7\textwidth}{!}{%
  \includegraphics{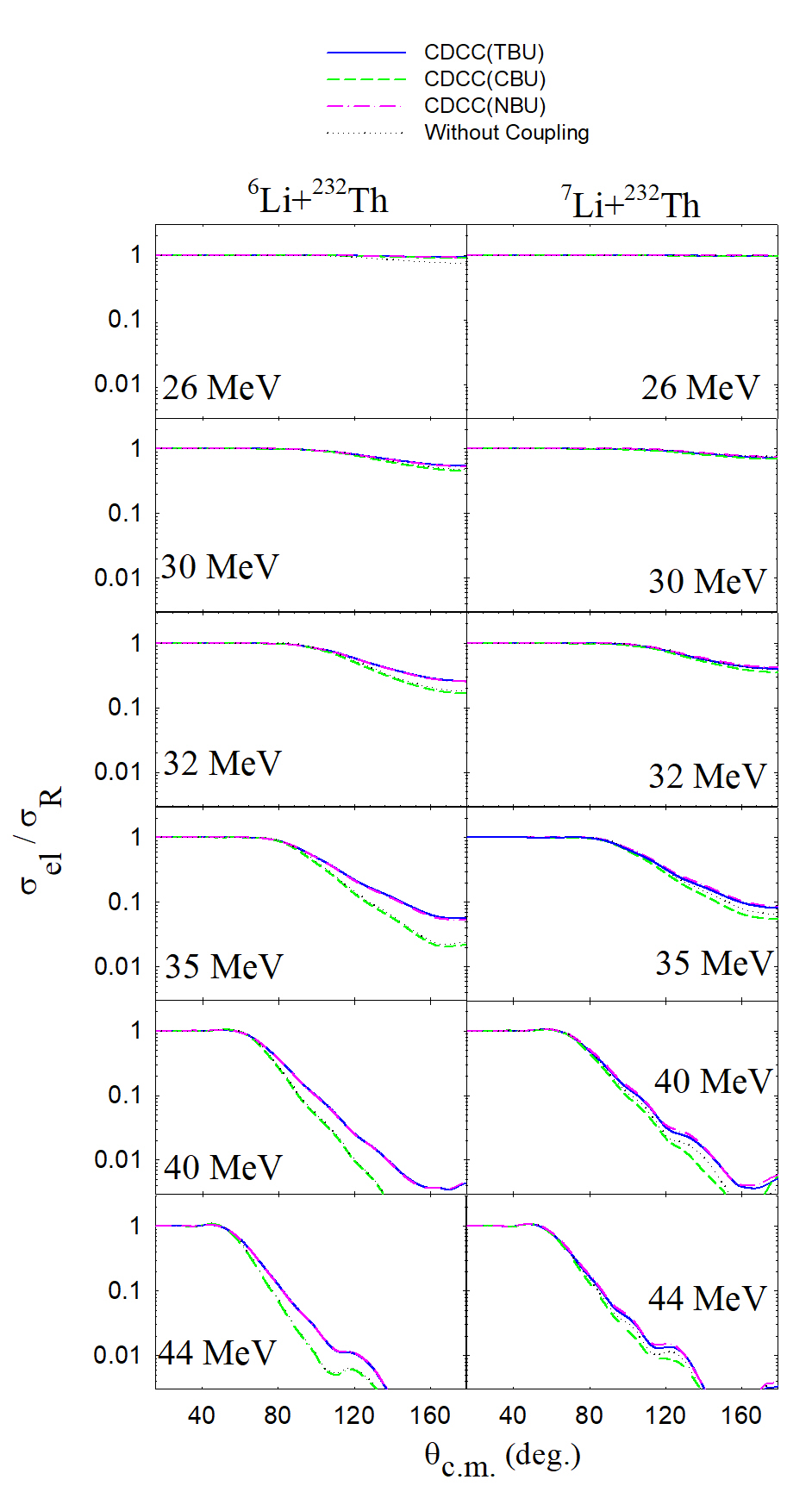}
}
\caption{(Color online)Elastic scattering angular distributions at 44, 40, 35, 32, 30, 26, and 24MeV for the $^{6,7}$Li+$^{232}$Th systems \cite{Dubey:2014ota}. The lines are the results of CDCC calculations. (refer to the text for detailed discussion).}
\label{fig:cn-67lith-img0.78.eps}       
\endcenter
\end{figure*}

\begin{figure*}
\center
\resizebox{0.99\textwidth}{!}{%
  \includegraphics{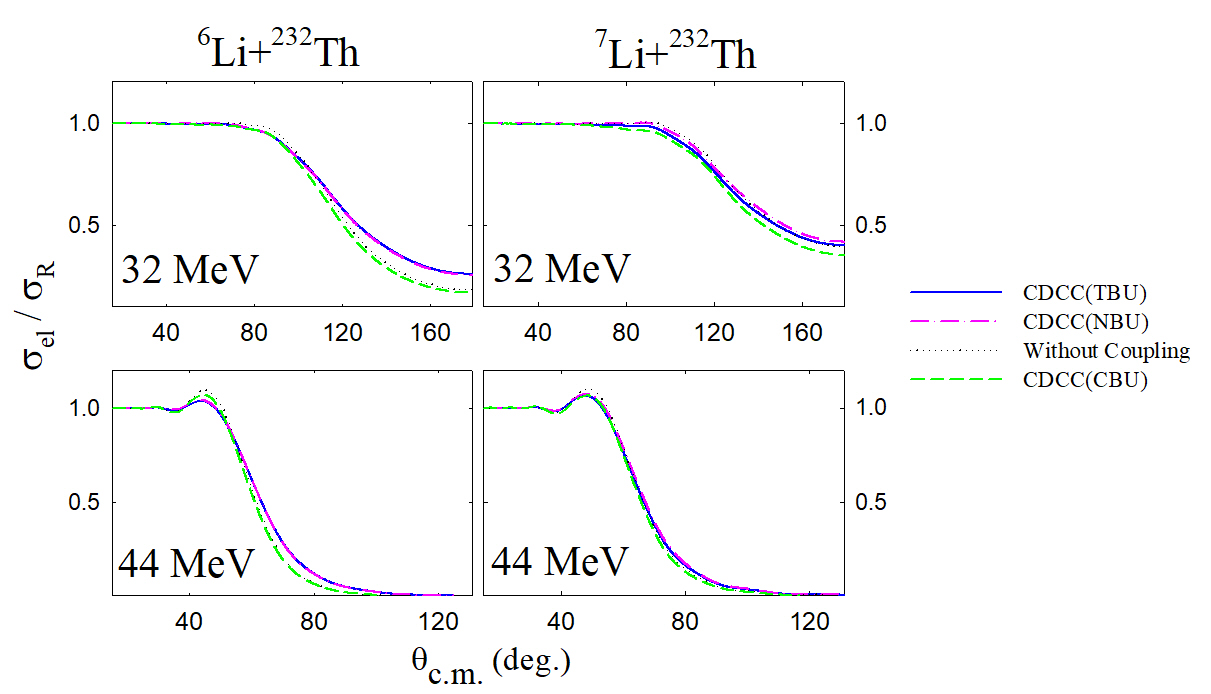}
}
\caption{(Color online)Elastic scattering angular distributions for $^{6,7}$Li+$^{232}$Th systems at the Coulomb barrier energy(26MeV) and above the Coulomb barrier(44MeV) are presented in linear scale. Experimental data \cite{Dubey:2014ota} are compared with the CDCC calculations (refer to the text for detailed discussion).}
\label{fig:c,n-linear-img0.78.eps}       
\endcenter
\end{figure*}
\begin{figure*}
\center
\resizebox{0.70\textwidth}{!}{%
  \includegraphics{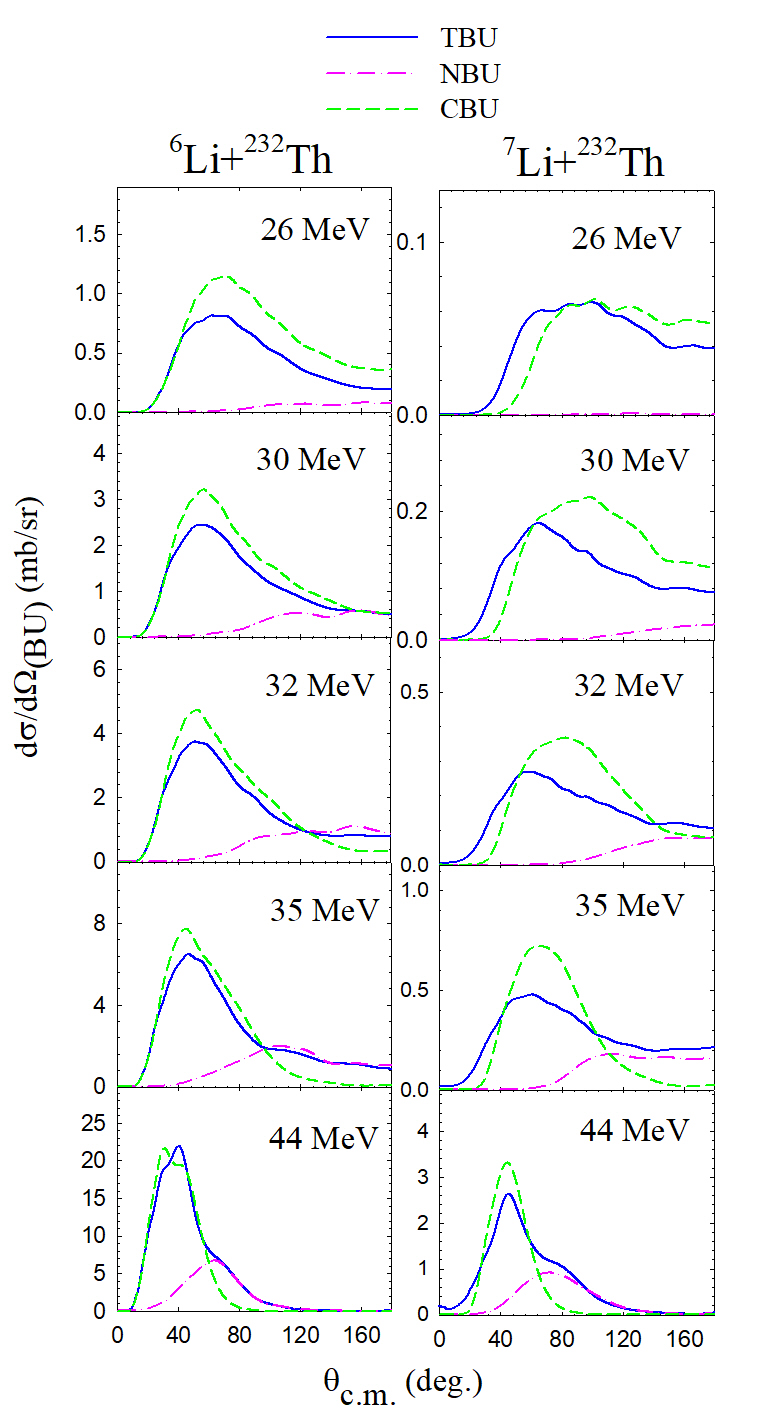}
}
\caption{(Color online)Comparison of Coulomb, nuclear, and total breakup cross-section angular distributions at 44, 35, 30, 32, 30, and 26MeV for the $^{6,7}$Li+$^{232}$Th systems (refer to the text for a detailed discussion).}
\label{fig:corrected-c,n-alpha-all-img0.78.eps}       
\endcenter
\end{figure*}

\begin{figure*}  
\center
\resizebox{0.7\textwidth}{!}{%
  \includegraphics{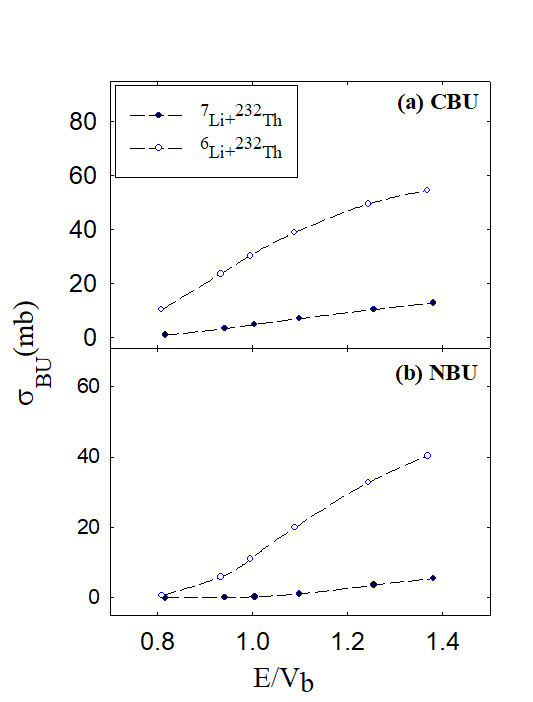}
}
\caption{(Color online)Integrated (a)Coulomb(CBU) (b)Nuclear(NBU) breakup cross-sections of $^{6,7}$Li projectiles on a $^{232}$Th target, derived from CDCC calculations at energies near the Coulomb barrier (refer to the text for a detailed discussion).}
\label{fig:5}       
\endcenter
\end{figure*}

\begin{figure*}
\center
\resizebox{0.7\textwidth}{!}{%
  \includegraphics{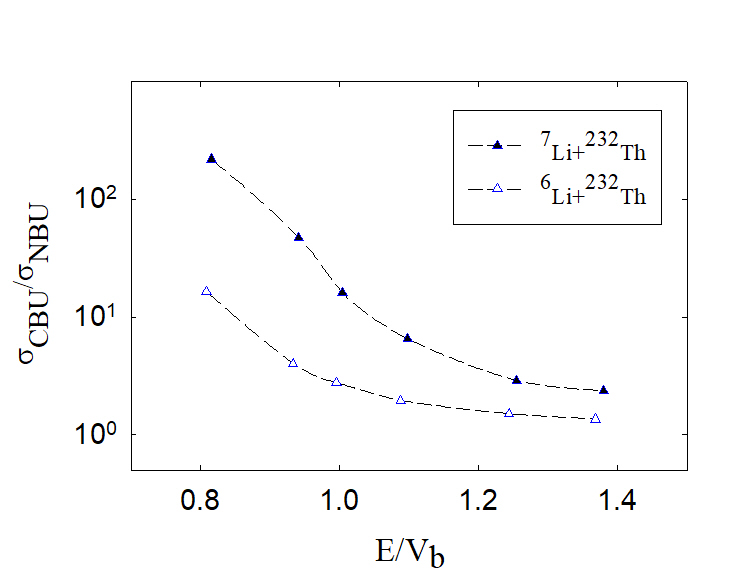}
}
\caption{(Color online)Ratio of integrated Coulomb and nuclear breakup cross-sections for $^{6,7}$Li projectiles with $^{232}$Th target, derived from CDCC calculations at energies around the Coulomb barrier (refer to the text for a detailed discussion).}
\label{fig:6}       
\endcenter
\end{figure*}

\begin{figure*}
\center
\resizebox{0.99\textwidth}{!}{%
  \includegraphics{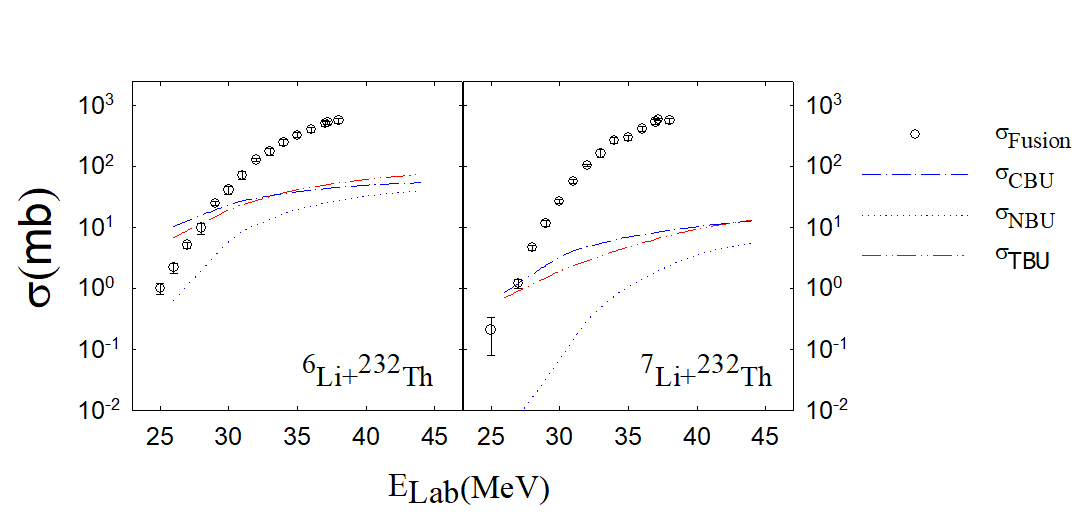}
}
\caption{(Color online)Comparison of experimental fusion cross-sections \cite{Freiesleben:1975zz} with the Coulomb, nuclear, and total breakup cross-sections from the present calculations for $^{6,7}$Li projectiles interacting with a $^{232}$Th target at energies near the Coulomb barrier (refer to the text for a detailed discussion).}
\label{fig:sig-react}       
\endcenter
\end{figure*}

\begin{table}
\center
\caption{Integrated breakup cross-sections, including total ($\sigma_{TBU}$), nuclear ($\sigma_{NBU}$), and Coulomb ($\sigma_{CBU}$) components, for a $^{6}$Li projectile interacting with a $^{232}$Th target, obtained from CDCC calculations at energies near the Coulomb barrier. The interference contributions are provided in the final column, with all cross-sections expressed in millibarns (mb).}
 \label{tab:3}
\begin{tabular}{lcccc}
\hline\noalign{\smallskip}
E$_{lab}$(MeV)& $\sigma_{CBU}$&  $\sigma_{NBU}$&$\sigma_{TBU}$& ${\sigma_{TBU}-\sigma_{NBU}} \over {\sigma_{CBU}}$\\
\noalign{\smallskip}\hline\noalign{\smallskip}
26&	10.40&	0.64&	6.88&	0.60\\
30&	23.58&	5.88&	19.45&	0.58\\
32&	30.33&	10.97&	27.72&	0.55\\
35&	38.89&	19.93&	41.73&	0.56\\
40&	49.50&  32.77&	61.67&	0.58\\
44&	54.58&	40.24&	74.21&	0.62\\
\noalign{\smallskip}\hline
\end{tabular}%
\endcenter
\end{table}

\begin{table}
\center
\caption{Integrated breakup cross-sections, including total ($\sigma_{TBU}$), nuclear ($\sigma_{NBU}$), and Coulomb ($\sigma_{CBU}$) components, for a $^{7}$Li projectile interacting with a $^{232}$Th target, obtained from CDCC calculations at energies near the Coulomb barrier. The interference contributions are provided in the final column, with all cross-sections expressed in millibarns (mb).}
 \label{tab:4}
\begin{tabular}{lcccc}
\hline\noalign{\smallskip}
E$_{lab}$(MeV)& $\sigma_{CBU}$&  $\sigma_{NBU}$&$\sigma_{TBU}$& ${\sigma_{TBU}-\sigma_{NBU}} \over {\sigma_{CBU}}$\\
\noalign{\smallskip}\hline\noalign{\smallskip}
26&	0.87	&0.004	&0.71	&0.81\\
30&	3.28&	0.07	&1.92	&0.56\\
32&	4.82	&0.3	&2.86	&0.53\\
35&	7.01	&1.08	&4.81	&0.53\\
40&	10.37	&3.6	&9.46	&0.57\\
44&	12.88	&5.46	&13.3	&0.61\\
\noalign{\smallskip}\hline
\end{tabular}%
\endcenter
\end{table}

Fig.~\ref{fig:cn-67lith-img0.78.eps} shows the angular distributions of the calculated elastic scattering normalized to Rutherford cross-sections at energies around the Coulomb barrier. In this figure, we are not showing the experimental data as pure elastic breakup cross-sections have been calculated. The continuous, dashed, dashed-dot, and dotted lines show total, Coulomb, Nuclear, and no coupling effects, respectively. In comparison to the $^{7}$Li+$^{232}$Th system, a noticeable breakup coupling effect is observed on the elastic scattering cross-section at energies above the Coulomb barrier for $^{6}$Li+$^{232}$Th. Additionally, it can be seen that the Coulomb-Nuclear Interference Peak (CNIP) is broader for $^{6}$Li than for $^{7}$Li, which aligns with expectations due to the lower breakup threshold of $^{6}$Li. This effect is more clearly depicted on a linear scale, as shown in Fig. \ref{fig:c,n-linear-img0.78.eps}. The influence of Coulomb couplings is evident at forward angles, while nuclear couplings become significant at backward angles.

Figure \ref{fig:corrected-c,n-alpha-all-img0.78.eps} shows the angular distribution of total, Coulomb, and nuclear breakup cross-sections at various energies around the barrier for the $^{6,7}$Li+$^{232}$Th systems. At energies above the Coulomb barrier, where the interacting nuclei are in closer proximity, nuclear breakup becomes more significant at larger angles, while Coulomb breakup remains dominant at forward angles. As the bombarding energy decreases, the transition angle from Coulomb to nuclear breakup shifts to higher angles.

As expected, Coulomb breakup becomes increasingly dominant at lower bombarding energies, reflecting the longer range of Coulomb interactions compared to nuclear interactions. This trend is also evident in the extracted breakup cross-section values presented in Tables \ref{tab:3} and \ref{tab:4}. Notably, near the barrier energies, Coulomb breakup cross-sections dominate the total breakup cross-sections. The total breakup cross-sections can be obtained by summing the nuclear and Coulomb breakup contributions, a relationship reflected in the calculated ratio ${\sigma_{TBU}-\sigma_{NBU}} \over {\sigma_{CBU}}$. The extracted values for this ratio, listed in the last columns of Tables \ref{tab:1} and \ref{tab:2}, deviate from unity across all energies for both $^{6,7}$Li+$^{232}$Th systems. This deviation suggests a destructive Coulomb-nuclear interference in the breakup process involving these projectiles and the heavy deformed $^{232}$Th target.

Figure \ref{fig:5} illustrates the calculated Coulomb and nuclear breakup cross-sections as functions of ${E}$/$V_{B}$ for the $^{6,7}$Li+$^{232}$Th systems. The results indicate that both Coulomb and nuclear breakup cross-sections are greater for $^{6}$Li than for $^{7}$Li. Additionally, the total breakup also remains higher across all energies for $^{6}$Li compared to $^{7}$Li, as shown in Tables \ref{tab:3} and \ref{tab:4}. This behavior is primarily attributed to the lower breakup-$\alpha$ threshold of $^{6}$Li (1.67 MeV). However, a careful analyses of Fig. \ref{fig:6} shows the relevant dipole response above breakup cross sections. This figure shows the ratio of Coulomb to nuclear breakup cross-sections and one can see that this is higher for $^{7}$Li than for $^{6}$Li. This ratio exceeds unity for both the $^{6,7}$Li+$^{232}$Th systems. However, as the incident energy increases, the ratio decreases, suggesting a diminishing relative influence of Coulomb forces compared to nuclear forces. This behavior can be related to the high Coulomb dipole response for $^{7}$Li. The dipole strengths $(BE1)$ is proportional to $|Z_1A_2 - Z_2A_1|$  where ${Z1,A1}$ and ${Z2,A2}$  are respectively the atomic and mass numbers of
the core and the valence particle in the single particle configuration. Thus the dipoles strengths areare 0 and 0.082 fm$^{4}$ e$^{2}$ for $^{6,7}$Li respectively ~\cite{Otomar:2015hqa}. A similar trend has been observed with spherical targets such as $^{208}$Pb when bombarded with $^{6,7}$Li projectiles. In contrast, for the lighter target $^{59}$Co, the ratio falls below unity for $^{7}$Li compared to $^{6}$Li \cite{Otomar:2009zz,Otomar:2012nr}.

Figure~\ref{fig:sig-react} presents a comparison of fusion cross-sections \cite{Freiesleben:1975zz} with total breakup (TBU), Coulomb breakup (CBU), and nuclear breakup (NBU) cross-sections for both $^{6,7}$Li+$^{232}$Th systems. For $^{7}$Li, the Coulomb breakup contribution remains higher than the total breakup cross-section. In contrast, for $^{6}$Li, the Coulomb breakup cross-sections increase as energy decreases, attributed to its low breakup threshold and the diminishing effect of the low-energy Coulomb dipole response. The experimentally measured fusion cross-sections are observed larger than the breakup components, except at the lowest energy points in both reactions. This indicates the presence of breakup reaction channels even at sub-barrier energies.

\section{Conclusions}
The role of Coulomb and nuclear couplings, as well as their interference effects, have been explored in the breakup processes of weakly bound projectiles $^{6}$Li and $^{7}$Li with the nonspherical $^{232}$Th target. Inclusion of $^{232}$Th inelastic states couplings exhibited strong hindrance below the barrier energies for $^{6,7}$Li+$^{232}$Th systems. The Continuum Discretized Coupled Channel (CDCC) method, which employs short-range interaction potentials for fragment-target interactions, provides a slightly better description of elastic scattering angular distributions compared to the method of scaling the imaginary potentials by a factor of 0.78 at around the Coulomb barrier energies. 

The standard Continuum Discretized Coupled Channels (CDCC) method, incorporating scaled SPP potentials ($0.78 \times W_{SPP}$), has been employed to determine the breakup cross sections at various energies for both $^{6,7}$Li+$^{232}$Th systems. The breakup cross-section has been effectively decomposed into its Coulomb and nuclear components, offering valuable insights into their respective contributions. The breakup cross sections for $^{6}$Li are found to be higher compared to those for $^{7}$Li. Above the barrier energies, Coulomb breakup dominates at forward angles, whereas nuclear breakup becomes more pronounced at larger angles.
For the $^{6}$Li+$^{232}$Th reaction, the Coulomb breakup cross-sections are larger than both the nuclear and total breakup cross-sections only at sub-barrier energies. In contrast, for the $^{7}$Li+$^{232}$Th reaction, the Coulomb breakup remains greater than the nuclear and total breakup cross-sections at all incident energies. This suggests the presence of destructive Coulomb-nuclear interference in the breakup processes involving weakly bound $^{6,7}$Li projectiles with the deformed $^{232}$Th target. A similar phenomenon was previously reported for a spherical target in Refs. ~\cite{Otomar:2009zz,Otomar:2012nr}. The $^{7}$Li+$^{232}$Th system exhibits a higher ratio of Coulomb to nuclear breakup cross-sections compared to the $^{6}$Li+$^{232}$Th, consistent with observations for a spherical target. These findings reveal the importance of incorporating detailed coupling effects to accurately characterize the reaction dynamics of weakly bound projectiles interacting with strongly deformed target nuclei.

\section*{Acknowledgement}
D. Patel acknowledges the financial support from CSIR-HRDG through a major research grant(Project file No. 03WS(002)/2023-24/EMR-II/ASPIRE). J. R. and J. L. acknowledge partial financial support from CNPq, FAPERJ, and INCT-FNA (Instituto Nacional de Ci\^ {e}ncia e Tecnologia- F\' isica Nuclear e Aplica\c {c}\~ {o}es), research project No. 464898/2014-5 and FAPERJ No. 13/2023.

\bibliography{RefGTMD.bib}
\bibliographystyle{hieeetr.bst}
\end{document}